\documentclass[submission,copyright,creativecommons]{eptcs}

\usepackage[T1]{fontenc}
\usepackage{xspace}
\usepackage[draft,inline,index,nomargin]{fixme}
\usepackage{graphicx, xspace, amssymb, latexsym, amscd}
\usepackage{amsmath, amssymb, amsbsy}
\usepackage{tikz}
\usepackage{moreverb}
\usepackage{ae}
\usepackage{color}
\definecolor{Blue}{rgb}{0.3,0.3,0.9}

\newcommand{\omitir}[1]{}

\newcommand{\Sect}[1]{Section~\ref{sec:#1}\xspace}
\newcommand{\Sects}[2]{Sections~\ref{sec:#1} and~\ref{sec:#2}\xspace}

\newcommand{\FormaLex}{\textsc{FormaLex}\xspace}
\newcommand{\cita}[1]{``\textit{#1}''\xspace}

\newcommand{\CL}{\ensuremath{\mathcal{CL}}\xspace}
\newcommand{\FL}{\ensuremath{\texttt{FL}}\xspace}

\newcommand{\until}{\,\, U \,}
\newcommand{\SF}{\ensuremath{\mathcal{F}}\xspace}
\newcommand{\SFsinP}{\ensuremath{\mathcal{F}_{\bar{P}}}\xspace}

\renewcommand{\O}[1]{\ensuremath{O(#1)}\xspace}
\renewcommand{\OE}[1]{\ensuremath{O^E(#1)}\xspace}
\newcommand{\OR}[2]{\ensuremath{O_{#1}(#2)}\xspace}

\newcommand{\F}[1]{\ensuremath{F(#1)}\xspace}
\newcommand{\FR}[2]{\ensuremath{F_{#1}(#2)}\xspace}

\renewcommand{\P}[1]{\ensuremath{P(#1)}\xspace}

\newcommand{\clause}[1]{\textbf{#1}\xspace}
\newcommand{\counter}[1]{\texttt{#1}\xspace}
\newcommand{\action}[1]{\texttt{#1}\xspace}

\newcommand{\actionoutput}[2]{\texttt{#1}.\textit{#2}\xspace}

\newcommand{\interval}[1]{\texttt{#1}\xspace}
\newcommand{\boolvar}[1]{\ensuremath{\textit{#1}}\xspace}
\newcommand{\DiamondI}[1]{\Diamond_\interval{#1}\xspace}
\newcommand{\HAP}{\texttt{HAPPENING}\xspace}
\newcommand{\NOTHAP}{\texttt{NOT\_HAPPENING}\xspace}
\newcommand{\JUSTHAP}{\texttt{JUST\_HAPPENED}\xspace}

\newenvironment{fl}
{\begin{small}\begin{tabbing}}
{\end{tabbing}\end{small}}

\newcommand{\Tr}[1]{\ensuremath{\mathit{Tr}(#1)}\xspace}

\providecommand{\event}{FLACOS 2011}

\title{A Software Tool for Legal Drafting\thanks{Partially supported by PICT 2007 532, UBACyT 20020090200116, 2002009020084 and 20020100200103.}}
\def\titlerunning{A Software Tool for Legal Drafting}

\def\authorrunning{D. Gorín, S. Mera \& F. Schapachnik}

\author{Daniel Gorín
	\institute{Departamento de Computaci\'on, FCEyN,\\
	Universidad de Buenos Aires, Argentina}
	\email{dgorin@dc.uba.ar}
\and
	Sergio Mera
	\institute{Departamento de Computaci\'on, FCEyN,\\
	Universidad de Buenos Aires, Argentina}
	\email{smera@dc.uba.ar}
\and
	Fernando Schapachnik
	\institute{Departamento de Computaci\'on, FCEyN,\\
	Universidad de Buenos Aires, Argentina}
	\email{fschapachnik@dc.uba.ar}
}

\begin{document}
\maketitle

\begin{abstract}
Although many attempts at automated aids for legal drafting have been made, they were based on the construction of a new tool, completely from scratch. This is at least curious, considering that a strong parallelism can be established between a normative document and a software specification: both describe what an entity should or should not do, can or cannot do.
In this article we compare normative documents and software specifications to find out their similarities and differences. The comparison shows that there are distinctive particularities, but they are restricted to a very specific subclass of normative propositions. The rest, we postulate, can be dealt with software tools. For such an enterprise the \FormaLex tool set was devised: an LTL-based language and companion tools that utilize model checking to find out normative incoherences in regulations, contracts and other legal documents. A feature-rich case study is analyzed with the presented tools.
\end{abstract}

\section{Introduction}
\label{sec:intro}

Although many attempts at automated aids for legal drafting have been made (e.g.,~\cite{tools_for_ald,CL,drcontract,mc_electronic_contracts,discordance_detection_in_regional_ordinance,corapi-norm}), they were based on the construction of a new tool, completely from scratch. This is at least curious, considering that a strong parallelism can be established between a normative document and a software specification: both describe what an entity should or should not do, can or cannot do. In the case of normative documents, it is a legal entity. In the case of software specifications, it is a piece of software. Is a software specification so different from a normative document? If it is not, why do not reuse the already existing machinery that can successfully analyze specifications?

In this article we compare normative documents and software specifications to find out their similarities and differences. The comparison shows that there are distinctive particularities, but they are restricted to a very specific subclass of normative propositions. The rest, we postulate, can be dealt with temporal model checkers. For such an enterprise the \FormaLex tool set was devised: an LTL-based language, \FL, and companion tools that utilize model checking to find out normative incoherences in legal documents.

\FL is based on the following key-concepts:
\begin{itemize}
\item It provides a \emph{background theory} to state matters about the real world, such as event precedence (e.g., sunrise before dawn), uniqueness (each person is born only once), etc., that would otherwise had to be accommodated into unnatural deontic rules. Said background theory is translated into an automaton that determines the class of models over which the rest of the language predicates. \Sect{translations} provides the details.
\item Deontic rules are translated into LTL, but the input language, that is, the way the original deontic rules are written, is preserved, so this information can be used to perform a set of analysis, at a meta-logical level. See \Sect{coherence_analysis} for details.
\item The combination of an automata-based formalism plus a logic that can refer to it is very powerful, and the software community knows it. \FL takes advantage of that to easily express properties that are generally difficult to pose in other formalisms, or lead to computational complexity problems. These features can be seen in \Sects{fl}{case_study}.
\end{itemize}

A comparison between specifications and legal documents is presented in \Sect{specs_vs_regulations}. \Sect{fl} highlights \FL, our LTL-based language, by describing its use to express otherwise hard-to-write properties, and \Sect{inner_workings} gives details of the tool's inner workings and formal semantics. A case study, where \FL is used to expose problems in a feature-rich university's regulation, is discussed in \Sect{case_study}.\footnote{Another \FL case study was presented in~\cite{GMS_JURIX2010c}, yet it was much smaller both in size and concepts.} \Sect{related_work} compares related work and \Sect{conclusions} concludes the article.

\section{Specifiable Regulations?}
\label{sec:specs_vs_regulations}

It is often understood that regulations can be abstractly represented using the three well-known deontic operators for obligation (O), prohibition (F) and permission (P). Specifications can be also thought of as using the same operators. \cita{The system must activate the brakes in no more than three seconds after the emergency stop button is pressed} is clearly an example of obligation. Prohibitions are also found: \cita{it is forbidden that the server sends unencrypted passwords through the wire}. Permissions are not that frequent, but also possible: \cita{if out of resources the system can drop requests until the processor is freed}.

Nevertheless, some types of statements found in legal documents are not common in software specifications. We divide them in two groups. The first one is composed of statements that have equivalents in specifications, but under a somehow different structure:

\paragraph{Contrary-To-Duty Obligations.}
Contrary-To-Duty (CTD) obligations are a way to model phrases like \cita{The agent is required to do action X. If she does not, then action Y should be performed}. The first sentence is called the \emph{obligation} and the last one the \emph{reparation}, and we denote that as $O_Y(X)$. Software specifications equivalents are conceivable, but the key difference is how to treat $O(X) \land O_Y(X)$. A Software Engineering perspective could read the formula as \cita{obligation to perform X and obligation to perform X, plus reparation Y in case of failure of X, equals obligation to perform X plus reparation Y}, considering that $O_Y(X)$ somehow supersedes plain $O(X)$. However, a legal point of view is that $O_Y(X)$ means that a behavior that does not satisfy X but performs Y is still legal, while the same behavior is illegal for the formula $O(X)$, that does not contemplate a reparation.

\paragraph{Amendments to Deal with Contradictions.}
A legal corpus might contain the formula F(kill), and then be modified to allow self-defense by the addition of P(kill in self-defense). It is hard to consider that such a corpus is contradictory, yet a software engineer will rather use the specification F(kill unless in self-defense) that does not entitle a logical contradiction.

\paragraph{Permissions.}
Besides their use as exceptions to prohibitions, what are permissions exactly? The question has been raised and analyzed many times before (e.g.,~\cite{expressive_conception_of_norms,permissions_and_obligations_in_hier_normative_systems,what_is_io_logic,ten_philosophical_problems_in_dl}), so let us here only say that in software a permission is little more than no-determinism, while in a normative system a permission is a much complex individual. It is worth noticing that the ability of a user of some application to use or not some functionality is not a permission, it is an obligation: the specification would probably say that the software is obliged to behave in a way or some other depending on the user's choices. Some phrases lend themselves to confusion: \cita{The user may print the displayed listing}, in the context of a software specification is just a simpler wording for stating that in order to comply with the specification the software is obliged to present the printing option, and, if chosen, print the listing.

\paragraph{Hierarchies.}
Legal corpora have hierarchies. For example a regional law might set a tax to \$10 while the national law sets the same tax to \$20. If national laws override regional ones in said normative system, the outcome is clearly \$20. A software engineer might be tempted to say that such system is equivalent to one that sets the tax to \$20. However, the real normative system permits that if the national tax-setting law is canceled, the tax is still set at \$10 by the regional one, while the software engineer model's does not.

\paragraph{Ontologies.}
In legal corpora ontologies are of common use. For instance, a law might set standards for animals such as pets, while there might be another, more specific for dogs, that might set different conditions. Although software engineers are familiar with inheritance and subclassing, it is not common to specify the requirements for a general class of actors and separately others, possibly contradicting the former, for a subclass. In a software specification they will be treated in an ad-hoc manner, for example, with a requirement for dogs and another for pets that are not dogs, thus avoiding the contradiction. \newline

There are other types of statements that we believe have no equivalent in software specifications:

\paragraph{Nesting of Deontic Operators.}
Nesting of deontic operators, as in obligation to obligate (or to forbid, or to permit). In normative propositions such as \cita{The judge is obliged to oblige the citizen to do X} there are two obligations (and two responsible parties): if the judge does not comply to oblige the citizen, she is to blame. If the judge complies but the citizen does not, then the citizen is at fault. The specification of a security system might use a similar phrase: \cita{The system is obliged to oblige other users to do Y}, with Y being something like \cita{not access each others private files}. In this case, if the system does not enforce Y, then the system is at fault. Also, if users fail to do Y, then the same system is also responsible. This type of predicates seem to be just a complex wording for only one obligation.

Care should be exercised when analyzing propositions where there is an apparent nesting of obligations, but can be rewritten without nesting. E.g., \cita{The voting system should oblige users to deposit the ballot in the case in less than one minute or else face prosecution}. Such a phrase has a very different meaning if found in the legal norm that regulates electronic vote, or in the software's specification. In the first case, it binds both developers and voters, while in the second only developers, as a software specification has no power over voters. In such a case, it can be rewritten as \cita{The voting system has to give one minute for the ballot to be deposited into the case. In case of timeout, prosecution actions should be initiated [i.e., by notifying officials].}.

\paragraph{Self-Referencing Modifications.}
Self-referencing modifications, as in \cita{Let Article X of Bill Y be modified to mandate that from now on such and such}. \emph{Self} here means that they modify the same normative systems that contains them. This should not be confused with any type of software compile-time or runtime configuration. In such cases the specification is still fixed and contemplates the different possible behaviors.\footnote{Although there are some prototype dynamic specification languages with self-referencing capabilities, they are still far from being used in the current state of the practice.}

\paragraph{Deontic-Conditional Validity.}
Deontic operators whose activating condition is the validity of another deontic operator. A typical example is a conditional over an obligation as in \cita{if at the time of the execution the agent were obligated to ... then she ...}. Software specifications might impose obligations based on the runtime operating conditions of the software, but they do not specify behaviour that is conditional to the \emph{runtime requirements}, if that term makes any sense at all. \newline

We found that the common denominator of the last group is considering the deontic operators as first-class operators, allowing for operators that take operators as parameters, check if they are active, and so on. However, we believe that if we consider only the legal documents that do not use such classes of predicates we can a) cover an important and varied amount of regulations that are common in the real world, and b) resort to the tools and technologies that can be used to analyze software specifications. The first group of predicates, we postulate, can be accommodated in such setting if treated properly.

We believe that this is good news for the deontic community, as it means that decades of effort in software-analysis tool building, optimizations and expressivity improvements can be leveraged, and there is no need to start from scratch and climb again the road from handling toy examples to real-size ones.

\section{The \FL Language}
\label{sec:fl}

Our starting point is that many contracts and regulations can be formally analyzed with tools originally aimed for software specifications. This allows for making the most out of existing tools.

\FL, introduced in~\cite{GMS_JURIX2010c}, is built on the following premises:
\begin{itemize}
\item It aims at finding \emph{coherence problems}, defined in a very pragmatic way: behaviours can not be permitted and forbidden, or obligated and forbidden, can not be plain mandatory or mandatory with CTDs, CTD reparations cannot be forbidden, etc. The complete list of covered topics is presented in~\Sect{coherence_analysis}.
\item The input language is divided into a \emph{background theory} and a set of rules. While the rules are LTL formulae with additional deontic operators aimed at capturing normative propositions, the background theory provides some simple constructs to describe the class of models over which the rules predicate.
\item Models are linear\footnote{It does not mean the branching alternatives are not present, each possible alternative is present in one of the considered models.}, and each one describes a possible \emph{legal} behaviour. That is, behaviours that do not comply with the normative rules are discarded.
\item If something is obligatory, then it must hold in every legal model at every possible state, and thus \O{\varphi} is interpreted as $\Box \varphi$. Prohibition of something is obligation to the contrary (\F{\varphi} $\equiv$ \O{\lnot\varphi}). The diamond operator works in the usual way, and $\Diamond \varphi$ looks ahead in the model for some state where $\varphi$ holds.
\item Contrary-To-Duty obligations are supported as \OR{\rho}{\varphi}, translated as $\Box (\lnot \varphi \to \rho)$. \FR{\rho}{\varphi} is interpreted as \OR{\rho}{\neg \varphi}. It is worth noticing that our encoding skips out most of the deontic logic paradoxes (see \cite[Sect.~6]{GMS_FLACOS2011-TR} for details).
\item Although based on translating to LTL, the input syntax is preserved to perform analysis at the meta-logical level.
\end{itemize}

Permission is thought of as absence of prohibition, but treated not as an operator that modifies the set of legal behaviours, but rather as a predicate that the legal models must satisfy. Otherwise, it is considered that the normative system under analysis (NSUA) has a \emph{coherence problem}: it states that something is permitted when it actually is not. If the user flags the permission as an exception to a prohibition, as in F(kill) and P(self-defense killing), the internal representation of the affected prohibition is changed to reflect that. In the example, to F(kill unless self-defense).\footnote{This can be done automatically for simple cases and requires manual intervention in others.}

The main component of the background theory is the \emph{action}. An action can be happening or not at any moment. In \FL an action is interpreted as a \emph{digital signal} that can be on or off for an arbitrary number of consecutive states. Actions can represent proper actions by the implicit agents (e.g., \clause{action} \action{DriveCar}) or non-controllable, external events (e.g., \clause{action} \action{CarCrash}). There is no explicit notion of a role performing an action, so if they are needed the subject must be encoded in the action. We plan to add this feature in the future.

Some requirements seem to be easy to express, like being licensed in order to drive a car. It would seem that it suffices to forbid the \action{DriveCar} action if there is no prior \action{GetLicense} action. But the easiness is only apparent, as individuals can not only get their license, but also lose it. If the \action{LoseLicense} action is also considered, establishing whether an individual is licensed or not by means of a pure formula amounts to ``parenthesis counting'', and that can be very challenging to write or plain impossible depending on the particular logic used. \FL incorporates the notion of \emph{intervals}, similar to the \emph{fluents} of~\cite{fltl}. An interval is delimited by beginning and ending actions, in a such a manner that there is no nesting nor closing of an already closed interval. In the automata, a propositional variable is set to true or false, indicating whether the interval is open (see \Sect{translations} for details). With intervals, the driving requirement can be posed as
\begin{fl}
\clause{interval} \interval{licensed} \clause{delimited by actions} \action{GetLicense}-\action{LoseLicense}
\end{fl}
\noindent and then simply stating \F{\lnot \boolvar{is\_licensed} \land \action{DriveCar}}.

Intervals can also be used to bound the occurrence of other actions:
\begin{fl}
\clause{interval} \interval{school\_time} \clause{delimited by actions} \action{CourseBegin}-\action{CourseEnds} \\
\clause{action} \action{TakeExam} \clause{occurs only in scope} \interval{school\_time}
\end{fl}

There is also the view of obligation not as something that must always hold, but rather as something that must be done, usually within some bound of time, sometimes called \emph{non-persistent obligation}. We can deal with such expressions in two ways. Either with the \OE{\varphi} operator that expresses an \emph{eventual} obligation, one that ceases to exists as soon as it is fulfilled, or, if time bounds are provided (e.g., \cita{You ought to return the borrowed books within school time}), by using intervals inside obligations: \O{\DiamondI{school\_time} \action{ReturnBook}}. Other interactions between obligations and deadlines are shown in~\cite{GMS_FLACOS2011-TR}.

As the background theory is translated into an automata, we can accommodate there (bounded) \emph{counting}, allowing for expressions like \O{\counter{bbc}>0 \to \Diamond \counter{bbc}=0}, where the integer counter \counter{bbc}, \emph{borrowed books count}, is handled by the automata incrementing it and decrementing it with every \action{BorrowBook} or \action{ReturnBook}, respectively. Such formula properly states that every borrowed book must be returned, whereas the simple \O{\action{BorrowBook} \to \Diamond \action{ReturnBook}} is satisfied by borrowing many and returning just one.

Also interesting are persistent obligations with one-time-each reparations, where violating the obligation once, or even performing the reparation should not free the subject from the obligation. An example of that is obligation to not cross red traffic lights, subject to a fee for each violation. Such CTD obligations are usually hard to express. For instance, \FR{\action{PayFine}}{\action{RedCrossing}}, says that red crossing is forbidden, except that a fine is payed immediately. If a diamond is added, as in \FR{\Diamond \action{PayFine}}{\action{RedCrossing}}, then many violations can be canceled with one payment.

In \FL the formula can be easily stated with the aid of a \counter{fines} counter that increments with \action{BeFined} and decrements with \action{PayFine}, and the following formulae: \FR{\action{BeFined}}{\action{RedCrossing}} (red crossing is forbidden, under the penalty of being fined), \O{\counter{fines}>0 \to \Diamond \action{PayFine}} (it is mandatory to eventually pay the fines).



\Sect{inner_workings} shows formal semantics, how the background theory is translated into automata, the handling of formulae and how the coherence checks are defined and performed.

\section{Semantics \& Inner Workings}
\label{sec:inner_workings}
\subsection{Background Theory}
\label{sec:translations}

\FL's background theory is translated into a Büchi automata network with additional code that controls state transitions and handles state variables. Each run of the automata defines a standard LTL model over which the rules formalized in the next section predicate. The tool can use SPIN~\cite{SPIN}, DiVinE~\cite{divine} or NuSMV~\cite{nusmv} as backends for model checking, but the encoding presented here will use an agnostic dialect.

Time modeling is discrete, considered as succession of states, some of which have a proper name to refer to timestamps of interest for the NSUA. In \FL an action is thought of as a \emph{digital signal} that can be on or off for an arbitrary number of consecutive states. Thus, each action is represented by an enumerated variable that covers the \HAP and \NOTHAP states. This is an easy way to model time density, as the net effect is that any event can happen while others are taking place.


One single automaton is responsible for controlling all the variables. It has a single state called \verb|running| and non-deterministically guarded self-transitions. Let's exemplify with the action \action{BorrowBook} (e.g., from the library).

\begin{small}
\begin{verbatimtab}
declare enum borrow_book = NOT_HAPPENING

running -> running 
	guard borrow_book = NOT_HAPPENING -> set borrow_book = HAPPENING;
	guard borrow_book = HAPPENING -> set borrow_book = NOT_HAPPENING;
\end{verbatimtab}
\end{small}

The automaton is therefore defined as follows:

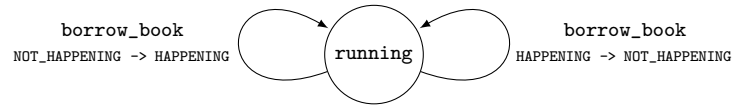
\begin{center}
\begin{scriptsize}
\begin{tikzpicture}[>=latex]
  \node (n1) at (0,0) [shape=circle,draw] {\texttt{running}} ;

  \draw (n1) edge [->, in=150, out=200, loop] node [left, text width=29mm] {\centering \texttt{borrow\_book} \\ \texttt{\tiny NOT\_HAPPENING -> HAPPENING}} (n1);

  \draw (n1) edge [->, in=30, out=340, loop] node [right, text width=29mm] {\centering \texttt{borrow\_book} \\ \texttt{\tiny HAPPENING -> NOT\_HAPPENING}} (n1);
\end{tikzpicture}
\end{scriptsize}
\end{center}

The automaton can switch the value of \verb|borrow_book| or leave it as it is, changing other variables. At the automaton level only one variable changes at a time, resembling the one input assumption of SCR~\cite{SCR}. As said before, at the normative level many actions can be taking place at the same time.

The encoding shown so far allows to easily refer to whenever actions are happening or not, but sometimes it is required to express that an action has happened completely (i.e., it has finished taking place) or just happened (i.e., has finished taking place in the current state). For instance \cita{account the loan after borrowing}, needs to refer to a moment when the \action{BorrowBook} action is not happening after having happened. To facilitate this, another state called \verb|just_happened|, of a type sometimes refered to as \emph{urgent} or \emph{committed}, is included in the automaton. The semantics of such type is that whenever an execution reaches one of these states, of all the available options, the automaton must execute a transition that leaves a committed state:

\begin{small}
\begin{verbatimtab}
running -> running 
	guard borrow_book = NOT_HAPPENING -> set borrow_book = HAPPENING;
running -> just_happened
	guard borrow_book = HAPPENING -> set borrow_book = JUST_HAPPENED;
just_happened -> running 
	guard borrow_book = JUST_HAPPENED -> set borrow_book = HAPPENING;
\end{verbatimtab}
\end{small}

With this new possible value for the state variable actions can be not happening for an arbitrary number of states, switch to \HAP, also for an arbitrary number of states, but before switching to \NOTHAP again they must spend one state as \JUSTHAP. Finished actions are easy to pose with this new state: in \O{\action{BorrowBook} \to \Diamond \action{AccountLoan}}. From the formula perspective, the terms \action{BorrowBook} and \action{AccountLoan} are translated to propositional variables whose value is \verb|borrow_book|=\JUSTHAP and \verb|account_loan| = \JUSTHAP respectively.

If actions have output values, as in: 
\begin{fl}
\clause{action} \action{BorrowBook} \clause{output values} \{ available, in\_house\_reading\_only, not\_available \}
\end{fl}
\noindent another enumerated variable \verb|borrow_book_output| is added to the automaton and the encoding turns into:

\begin{small}
\begin{verbatimtab}
...
running -> just_happened
	guard borrow_book = HAPPENING -> set borrow_book = JUST_HAPPENED,
					borrow_book_output = AVAILABLE;
	guard borrow_book = HAPPENING -> set borrow_book = JUST_HAPPENED,
					borrow_book_output = IN_HOUSE_READING_ONLY;
	guard borrow_book = HAPPENING -> set borrow_book = JUST_HAPPENED,
					borrow_book_output = NOT_AVAILABLE;
...
\end{verbatimtab}
\end{small}

So the output is set non-deterministically to any of the possible values and it is retained until the next setting. Similarly, if the action has extra guards, they are added to the \verb|guard| of the transition. 

As we mentioned before, intervals are another important feature of the language. Let's suppose books can only be borrowed during the academic year:

\begin{fl}
\clause{interval} \interval{academic\_year} \clause{delimited by actions} \action{BeginYear}-\action{EndYear} \\
\clause{action} \action{BorrowBook} \clause{occurs only inside} \interval{academic\_year}
\end{fl}

A boolean variable, \verb|academic_year_opened| is added to the automaton, and is set by the transitions that represent the respective actions. Also, it is added as a guard, so no \action{BeginYear} happens inside an academic year and no \action{EndYear} happens outside one. Similarly, it is added as a guard for \action{BorrowBook}.


Expressivity-wise, counters are a very powerful feature: they are basically a non-negative integer variable with actions that either increment, decrement or reset their value. The implementation is straightforward: an integer variable is added to the automaton, and it is manipulated in the respective transitions.

Temporal actions are a way to implement timers. The \clause{temporal actions} $ta_1, \ldots, ta_n$ clause declares a sequence of time points that follow the specified order and let an arbitrary number of actions happen between them. The implementation uses another synchronizing automaton with one state representing each $ta_i$ and others representing the time intervals after $ta_i$ and before $ta_{i+1}$.

\subsection{\FL's Syntax and Semantics}
\label{sec:syntax_and_semantics}

\newcommand{\prop}{\textsc{props}\xspace}
\newcommand{\inter}{\textsc{intervals}\xspace}
\newcommand{\forms}{\textsc{forms}\xspace}
\newcommand{\innerforms}{\textsc{inner\_forms}\xspace}

We present here a formal definition for the rule-stating part of \FL.\footnote{This reduced presentation does not allow for nesting of deontic operators, a restriction only introduced to save space in this article. For the same reason we ommit the repared version of the $O^E$ operator.} Although allowed in the input language, general terms like $\action{bcc} > 0$, \action{action.value}, etc.~are abstracted away as propositional symbols in the presented syntax. There is no need to model them explicitly since they can be thought of as encoded in propositional values that later the model handles in the proper way. The same happens with actions \action{a}, which are abstracted as the proposition \verb|a=|\JUSTHAP.

\paragraph{Syntax.} Let \prop be a countable infinite set of symbols, $\inter \subseteq (\prop \times \prop)$ a set of intervals, and \forms the set of \FL formulae in the signature $\langle\prop, \inter\rangle$ defined as

\begin{center}
$\innerforms ::= \top \mid \bot \mid p \mid \lnot\varphi \mid \varphi_1 \land \varphi_2 \mid \Diamond \varphi \mid \DiamondI{i} \varphi$
$\forms ::= O(\varphi) \mid F(\varphi) \mid \OR{\rho}{\varphi} \mid \FR{\rho}{\varphi} \mid \OE{\varphi} \mid P(\varphi),$
\end{center}

\noindent where $p \in \prop$, $\varphi, \rho, \varphi_1, \varphi_2 \in \innerforms$ and $\interval{i} \in \inter$. We usually work with (finite) sets of \forms when specifying a NSUA, so conjunction between formulae in \forms does not need to be formally defined. We will usually write one formula below another, implicitly defining a conjunction between them. Some operators could have been defined in terms of others, but we intentionally define all of them at this level since our tool considers them differently for coherence analysis (see \Sect{coherence_analysis} for more details).

\paragraph{Semantics.} We give \FL semantics by providing a translation $\mathit{Tr}$ from \FL into classic LTL\footnote{That is, the basic version of LTL with the standard boolean connectives plus the \textit{until} operator (from which the diamond is defined).}, as both work over the same class of models. Let \SF be a set of \FL formulae  whose intended meaning is defining the set of legal models for the NSUA. Recall that LTL models are linear infinite structures that represent possible runs on the automata defined by the background theory. We first split permissions from the rest and define the $\mathit{Tr}$ domain as $\SFsinP = \{\varphi \mid \varphi \in \SF \text{ such that $\varphi$ is not of the}$ $\text{form $P(\psi)$}\}$. $\mathit{Tr}$ acts as the identity for the \innerforms constructions not explicitly specified and it is asumed that the target LTL signature has the implicitly defined propositions involved in the translation (like $i\_opened$ for each interval $i$).
\begin{center}
$\begin{array}{rclrcl}
\Tr{\DiamondI{i} \varphi} & = & \interval{i}\_\mathit{opened} \to (\interval{i}\_\mathit{opened} \until \Tr{\varphi}) \hspace{10mm} & \Tr{O(\varphi)} & = & \Box \Tr{\varphi} \\
\Tr{F(\varphi)} & = & \Box \lnot \Tr{\varphi} & \Tr{\OR{\rho}{\varphi}} & = & \Box(\lnot \Tr{\varphi} \to \Tr{\rho}) \\
\Tr{\FR{\rho}{\varphi}} & = & \Box (\Tr{\varphi} \to \Tr{\rho}) & \Tr{\OE{\varphi}} & = & \Diamond \Tr{\varphi} \\
\end{array}$
\end{center}
\noindent $\mathit{Tr}$ can be lifted to take a set of \FL formulae and return the set of their translations.

\newcommand{\CASF}{\ensuremath{\mathcal{C}_A^\SF}\xspace}

Let's now consider an automaton $A$ defined by the background theory and the class of models $\mathcal{C}_{A}$ that represents all possible runs on $A$. Let \SF be the set of \FL formulae that encode the NSUA. The class of legal models defined by \SF over $\mathcal{C}_{A}$ is defined as $$\CASF = \{\mathcal{M} \in \mathcal{C}_A \mid \mathcal{M} \models \Tr{\SFsinP}\}.$$ That is, every legal model must satisfy the obligations and prohibitions specified by \SF.

But what about permissions? Permissions are actually a \emph{check} that must be performed on \CASF to ensure coherence. The condition that \CASF must fulfill is the following: for every $\varphi$ of the form $P(\psi)$ in \SF there must be a model $\mathcal{M}$ in \CASF such that $\mathcal{M} \models  \Tr{\psi}$. I.e., if something is permitted then the rest of the NSUA does not prevent it from happening.

We are going to expand the concept of coherence in the next section by analyzing other cases of interest. 

\subsection{Analyzing Coherence}
\label{sec:coherence_analysis}

A difficult topic in deontic logic is the concept of \emph{coherence} of a normative system: whenever the set of rules is ``contradictory'' in any sense. As stated in the literature (e.g., \cite{ten_philosophical_problems_in_dl}), the problem cannot be simply reduced to logical consistency. We take a pragmatic approach where a normative system is not coherent if it has any of a list of problems.

While some of them are straightforward to check, others require more sophistication. Let's see an example of each class. To fix notation, $\varphi \# \psi$ means that $\varphi$ is incompatible with $\psi$ and the following equivalences hold: \O{\varphi} = \OR{\bot}{\varphi}, \FR{\rho}{\varphi} = \OR{\rho}{\lnot \varphi}.

To check that there are no forbidden obligations (i.e., that there is no pair of rules \O{\varphi} and \O{\psi} such that $\varphi \# \psi$), we need to check that there is at least one possible legal behaviour that satisfies both the background theory and the complete set of formulae. To do that, all the rules $r_i$ are conjuncted into $\Phi = \bigwedge \Tr{r_i}$, and both the automata and $\lnot\Phi$ are fed to the model checker. If $\lnot\Phi$ is not satisfiable the model checker will output a counter example trace, $\tau$. $\tau$ satisfies the negation of $\lnot\Phi$ so it is the legal behaviour we were looking for. Should $\lnot\Phi$ be satisfiable, that means that $\Phi$ is not, so a backtracking-type of procedure should be started to find the ``guilty'' rules. How to improve this procedure is an active area of research.

We are also interested in checking that there are no ``contradicting obligations'': rules \OR{\rho}{\varphi} and \OR{\rho'}{\psi} with $\varphi \# \psi$ even if it is not the case that $\rho \# \rho'$. That is, incompatible obligations, but with compatible reparations. If that were the case, there would be a legal behaviour: doing the reparations $\rho$ and $\rho'$, yet it makes no sense that the primary obligations $\varphi$ and $\psi$ are impossible to comply with.

To check for that, we build $\Phi'$ as $\bigwedge \Tr{\O{\varphi_i}}$ for all the rules $r_i = \OR{\rho_i}{\varphi_i}$. Then the automata and $\lnot\Phi'$ are fed to the model checker as before. If $\lnot\Phi'$ is satisfiable there is no way to comply with all the obligations, leaving aside the possible reparations. If $\lnot\Phi'$ is not, then, as before, the $\tau'$ counter example trace is a possible way of complying with all the primary obligations.

It should be noted that this last check is one of the analysis were preservation of input syntax is important and the translation of repared obligations is not done directly.

The following conditions also violate coherence:

\begin{itemize}
\item \emph{Forbidden reparations}, such as $O_{\rho}(\varphi)$ and $F(\rho)$. In that case the reparation is only nominal, as it is indeed forbidden. $O_{\rho}(\psi)$ and $O(\psi)$ with $\rho \# \psi$ is another case of the same problem.
\item \emph{Obligations with conflicting reparations}. If $\rho \# \rho'$ and $O_{\rho}(\varphi)$ and $O_{\rho'}(\varphi)$ is found then there is a contradiction in how the obligation $\varphi$ could be repaired. Note that this does not mean that there is no legal behaviour, as respecting $\varphi$ is always allowed.
\item \emph{Impossible permissions}. Whatever was said to be permitted should be possible as was explained in \Sects{fl}{syntax_and_semantics}.
\item \emph{Unrealizable background theory}. The background theory should not generate an empty set of traces, which would mean it is, by itself, logically inconsistent.
\end{itemize}

\section{Case Study}
\label{sec:case_study}

To show \FL at work we analyze some excerpts from a university regulation. This case study focuses on conflicts that can arise from students being able to be also teachers. Although fictional, the inspiration is real. The case study features the use of actions, intervals, counters, obligations, prohibitions, permissions and many forms of coherence analysis.

The analyzed excerpt is the following:
\begin{small}
\begin{enumerate}
\item Chapter 1, Students.
 \begin{enumerate}
 \item Every individual that has enrolled for a career and has not yet graduated from it is considered to be a student.
 \item Students should respect each other. Major disciplinary faults are punished by forbidding the entering to university premises for one year after the fault.
 \item Students have the following rights: ..., participate in research activities, ... \label{perm_students_research}
 \end{enumerate}
\item Chapter 2, Teachers.
 \begin{enumerate}
 \item There are three teaching categories: c1) Undergraduate Teaching Assistant (aka UTA), c2) Teaching Assistant and c3) Professor.
 \item To become a teacher, aspirants must apply when the selection opens. The selection will be made based on the following criteria for each category: [omitted, not relevant to the case study]
 \item To apply for the UTA category, aspirants must be students at the time of the selection.\footnote{The UTA category position lasts one year in the real case. This particular spelling of the norm was chosen because it is desired that only students fill this position, yet allow them to keep the job if they graduate after the selection.}\label{art:student-teacher}
 \item Teachers must perform their duties, starting 30 days after the selection ends.
 \item Working from home is allowed, but teachers must spend at least one day a week in the premises of the university.
 \end{enumerate}
\item Chapter 3, Research.
 \begin{enumerate}
  \item Research activities can only be pursued by members of approved research groups.
  \item Research groups are conformed by accredited professors or teaching assistants.
 \end{enumerate}
\item Chapter 4, University Library.
 \begin{enumerate}
 \item Every borrowed book should be returned by the end of the month.\footnote{The more realistic requirement of returning within days is also possible but more involved, thus the simpler version is preferred for space reasons.} \label{art:borrow-return}
 \item Students and teachers are subject to a fine for not returning books in time. \label{art:fine}
 \item As students are generally on a budget, their fine should be low. \label{art:fine-students}
 \item Teachers should be an example of conduct, thus their fine should be strictly higher that the students' one. \label{art:fine-teachers}
 \end{enumerate}
\end{enumerate}
\end{small}

Let's model the student's chapter first. To avoid clutter some abbreviations will be used, such as not declaring actions that are used to bound intervals if they do not take any extra parameter, as it is the case for declaring what a student is. 

\begin{fl}
\clause{interval} \interval{student} \clause{delimited by actions} \action{Enroll}-\action{Graduate}
\end{fl}

Regarding discipline, they should not commit disciplinary faults or be banned from entering the premises for one year. To model that we will define two intervals: one that spans from the fault to one year after, and another that accounts for being inside the building.
\begin{fl}
\clause{interval} \interval{ban} \clause{delimited by actions} \action{CommittFault}-\action{OneYearPassed} \\
\clause{interval} \interval{inside\_building} \clause{delimited by actions} \action{Enter}-\action{Exit}
\end{fl}
\setlength\abovedisplayskip{9pt plus 1pt minus 7pt}
\setlength\belowdisplayskip{9pt plus 1pt minus 7pt}
And stipulate the prohibition:
\begin{equation}
\FR{\DiamondI{ban} \lnot \boolvar{is\_inside\_building}}{\action{CommitFault}} \label{form:fault}
\end{equation}
meaning that the fault should not occur but if it does, during the ban period the student cannot be inside the building (\boolvar{is\_inside\_building} is a boolean variable made true between the bounding actions of the interval).

Article \ref{perm_students_research} permits students to do research, in what can be thought of as a case of antithetical permission~\cite{DBLP:journals/japll/Stolpe10}: a permission set to invalidate future prohibitions.

\begin{equation}
\P{\boolvar{is\_student} \land \action{DoesResearch}} \label{form:perm_students_research}
\end{equation}

Now let's focus on the teachers' selection process. There is the selection interval and its possible outcomes: being elected in any of the categories, or not being elected at all.
\begin{fl}
\clause{action} \action{ElectWinners} \clause{output values} \{ teacher\_c1, teacher\_c2, teacher\_c3, no\_teacher \} \\
\clause{interval} \interval{selection} \clause{delimited by actions} \action{OpenSelection}-\action{ElectWinners} \\
\clause{action} \action{Apply} \clause{only occurs in scope} \interval{selection} \\
\end{fl}

For simplicity let's only model the requirements for the c1 category (UTAs) of still being a student:
\begin{equation}
\O{\DiamondI{selection}(\action{Apply} \to \boolvar{is\_student})} \label{form:student-teacher}
\end{equation}

Teachers also have duties, that must start 30 days after the election:
\begin{fl}
\clause{interval} \interval{grace\_period} \clause{delimited by actions} \action{ElectWinners}-\action{30DaysAfter} \\
\clause{interval} \interval{on\_duty} \clause{delimited by actions} \action{30DaysAfter}-\action{+inf} \\
\clause{interval} \interval{week} \clause{delimited by actions} \action{StartWeek}-\action{EndWeek} \clause{occurs only in scope} \interval{on\_duty} \clause{repeatedly}
\end{fl}

It is mandatory to have a weekly visit:
\begin{equation}
\O{\boolvar{is\_teacher} \to \DiamondI{week} \action{Enter}} \label{form:weekly_visit}
\end{equation}
\noindent with \boolvar{is\_teacher} defined as\footnotemark
\begin{fl}
\clause{macro} $\boolvar{is\_teacher} =~$\=\action{Apply} $\land$\\
\> (\actionoutput{ElectWinners}{teacher\_c1} $\lor$ \actionoutput{ElectWinners}{teacher\_c2} $\lor$ \actionoutput{ElectWinners}{teacher\_c3})
\end{fl}
\footnotetext{Note that although the name of the \action{Apply} action is in the present tense, because of the translation, it is easier read if it thought of as written in the past tense. This reflects the fact that the intended semantics is to check if the action has occurred at some point in the past. The same happens with formulae \ref{form:research_groups1}, \ref{form:research_groups2} and \ref{form:fix1}.}

Regarding Chapter 3, the restriction of research activities to members of research groups is:
\begin{equation}
\F{\lnot \action{JoinResearchGroup} \land \action{DoesResearch}} \label{form:research_groups1}
\end{equation}
\noindent while the requirements of being TA or professor to belong to a research group is:
\begin{equation}
\F{\lnot (\actionoutput{ElectWinners}{teacher\_c2} \lor \actionoutput{ElectWinners}{teacher\_c3}) \land \action{JoinResearchGroup}} \label{form:research_groups2}
\end{equation}

Then there is the borrowing of books, similar to both students and teachers, that requires the \counter{bbc} (borrowed books count) counter and signaling months (it should be noted that the exact duration of a month is not important and it is thus abstracted away):
\begin{fl}
\clause{counter} \counter{bbc} \=\clause{increases with action} \action{BorrowBook} \clause{decreases with action} \action{ReturnBook} \\
\clause{interval} \interval{month} \clause{delimited by actions} \action{MonthBegin}-\action{MonthEnd} \clause{repeatedly}
\end{fl}

Although articles \ref{art:fine-students} and \ref{art:fine-teachers} are mainly motivational, there is one prescriptive consequence, even at the level of abstraction we are using -- fines should be different: \action{StudentFine} \# \action{TeacherFine}.

Article \ref{art:fine} is a CTD to \ref{art:borrow-return}, so we encode them as:
\begin{align}
\OR{\action{StudentFine}}{\boolvar{is\_student} \to \DiamondI{month}(\counter{bbc}>0 \to \DiamondI{month} \counter{bbc}=0)} \label{form:library-students} \\
\OR{\action{TeacherFine}}{\boolvar{is\_teacher} \to \DiamondI{month}(\counter{bbc}>0 \to \DiamondI{month} \counter{bbc}=0)} \label{form:library-teachers}
\end{align}

When analyzed by \FormaLex three coherence problems are pointed out. First, the reparation for not returning borrowed books is troublesome for teachers that are also students, and the tool signals a case of \emph{conflicting reparations}, as there are traces where the implicit agent is indeed a teacher and a student. Looking at the NSUA nothing impedes students to become teachers, quite the contrary, as there is a UTA category.

What type of fine should a UTA be subject to? Whatever conclusion can be obtained by looking at the motivational articles \ref{art:fine-students} and \ref{art:fine-teachers} is disputable, as it is both true that UTAs should be an example of conduct because they are teachers, and that they are also on a budget, because UTAs' salaries are symbolic. As the fining would probably be decided by a library's clerk, there is high risk of different solutions applied to identical cases. It would be much better if this case could be decided by the norm-givers, and that is the sense of the warning.

The second problem is more involved and related to the weekly visit rule. The tool detects that the reparation of the rule that forbids discipline faults (\ref{form:fault}) contradicts the obligation to the weekly visit (\ref{form:weekly_visit}). Indeed, there is the possibility of a student committing a fault, thus being banned to enter the premises, applying for a teaching position, then being elected as UTA and not being able to comply with his weekly visit duty.

A possible solution is to forbid the application of punished students as in:
\begin{fl}
\clause{action} \action{Apply} \clause{only occurs in scope} \interval{selection} \clause{requires that} $\lnot$\action{CommitFault}
\end{fl}

However, the problem persists, as the student can now apply (not yet being faulty), commit the fault, then being selected, to the same effect. A better solution would be to restrict the \action{ElectWinners} action so only non faulty students can become teachers:
\begin{align}
\F{\action{CommitFault} \land \actionoutput{ElectWinners}{teacher\_c1}} \label{form:fix1}
\end{align}

This formula states that it is forbidden to commit the fault and to be elected UTA. Observe that if the fault is commited before becoming a teacher, then this formula forbids the election, which is the primary intention of it. On the other hand, if the fault is commited after becoming UTA, then again the reparation of the rule that forbids discipline faults (\ref{form:fault}) contradicts the obligation to the weekly visit (\ref{form:weekly_visit}). This situation is also noticed by the tool and notified to the user as a warning. If the user considers that it should be possible to repair the fault even under these conditions, then she must introduce appropriate modifications in the NSUA to envisage this situation. If, on the other hand, she thinks that it is correct to tighten the rules for teachers so they should not have a way to repair their faults, then the warning can be ignored.

The third problem is the collision of allowing research only to TAs and professors (rules \ref{form:research_groups1} and \ref{form:research_groups2}) with the permission for students by rule \ref{form:perm_students_research}. The fix is either the removal of the permission or the inclusion of at least UTAs into research groups.

There is another interesting aspect of this NSUA. Let's assume somebody proposes a different writing for article \ref{art:student-teacher}, that is supposedly more faithful to the spirit of not letting graduates fill the UTA category. She proposes defining what a graduate is:
\begin{fl}
\clause{interval} \interval{graduate} \clause{delimited by actions} \action{Graduate}-\action{+inf}
\end{fl}
\noindent and plain forbidding the application of graduates. When queried about the validity of this new writing:

? \F{\DiamondI{selection} (\action{Apply} \land \boolvar{is\_graduate})}

\noindent the tool responds that the prohibition does not hold as there is a trace where a student graduates and then enrolls again (say, for a different career) before applying. That perfectly satisfies the requirement that during the selection process applications must be done by students (\ref{form:student-teacher}), as the implicit agent is \emph{also} a student. It means that the phrasing of the rule (\ref{form:student-teacher}) does not comply with the intended normative effects: restrict the UTA category to non-graduates; so the proposed alternate writing is actually the correct one. This ``bug'' is extracted from a real university's regulation.

An interesting remark is that although the lack of roles in \FL can be seen as a drawback, it turned useful in this case. Should roles had been available, probably many of the above mentioned bugs would have been concealed, including the real one.

\section{Related Work}
\label{sec:related_work}

The idea of using a temporal logic for deontic purposes is not new. It can be traced to \cite{ntl_journal,meyer1998role,DBLP:conf/icail/GovernatoriRS05,SED-LTL2006,DBLP:conf/deon/FrenchMR10,DBLP:conf/deon/BroersenDDM04,piolle2010dyadic}, among others who use some type of temporal-deontic logic. However, as far as we know they provide neither tool support nor a translation into a standard tool, and thus are not directly comparable to our work which is heavily tool-biased.

\cite{discordance_detection_in_regional_ordinance} deals with automated conflict detection in norms by using a tool that supports ontologies and translates normative propositions into a Prolog program, but the analysis is restricted to logical contradiction. More similar to our approach of analyzing contracts and regulations for coherence problems are BCL~\cite{drcontract} and \CL~\cite{CL}. BCL is a contract specification language that is meant for monitoring, allows to build executable versions, can detect conflicts among rules off-line and provides features like clause normalization. However, it lacks support for temporal reasoning and background theories.

\CL is another logical language based on dynamic logic that treats deontic operators as first-class citizens. It is based on an ad-hoc tool and it neither uses background theories, nor discusses how to deal with some limitations of expressivity: for instance, in dynamic logic approaches it is easy to say that if a book is borrowed ($b$), a book should be returned ($r$) as [b]O(r), but such rule matches the borrowing of multiple books and the returning of just one; the correct version is pretty involved or plain impossible depending on the particular logic.



\section{Conclusions}
\label{sec:conclusions}

Starting from the premise that normative systems are very similar to software specifications we propose a language and related tool set to analyze the former with tools designed for the latter. Besides the similarities, the decision is based on avoiding to build tools from scratch: model checkers have decades of effort in tool building, optimizations and expressivity improvements.

In this article we analyzed a feature-rich, real-life inspired, case study. Although fictional, we believe it shows the power of both the tool and its underlying definition of coherence to spot defects that are not self-evident. Our next step is dealing with a 100\% real case study to investigate the payoff of having to logically encode the NSUA vs.~the severity of the defects found.

\bibliographystyle{eptcs}
\bibliography{formalex}

\end{document}